\documentclass{achemso}
\usepackage[version=3]{mhchem} 
\usepackage{booktabs,braket,mathptmx,xcolor,dcolumn,graphicx,bm,graphics,caption,multirow,array,paralist,rotating}
\usepackage[normalem]{ulem}
\usepackage[shortlabels]{enumitem}
\usepackage[at]{easylist}
\SectionNumbersOn
\usepackage{float,enumitem}
\usepackage{amsmath, amsfonts, amssymb, amsthm, xparse, calrsfs}
\usepackage{bbold}
\usepackage[UKenglish]{babel}
\usepackage[UKenglish]{isodate}
\usepackage{physics}
\usepackage{bm}
\usepackage{parskip}
\usepackage{esvect}
\usepackage[T1]{fontenc}
\usepackage{setspace}
\usepackage{subfig}
\usepackage{svg}
\cleanlookdateon
\usepackage[labelfont=bf]{caption}
\usepackage[justification=centering]{caption}
\usepackage{multicol}
\usepackage{enumitem}
\usepackage{enumitem}
\usepackage{color}
\usepackage{wrapfig}
\usepackage{newtxtext,newtxmath}
\newcommand{\kJmol}{$\mathrm{kJ \ mol^{-1}}$}

\newcolumntype{H}{>{\setbox0=\hbox\bgroup}c<{\egroup}@{}}
\newcolumntype{d}{D{.}{.}{-1}}

\usepackage[hidelinks]{hyperref}
\hypersetup{
  unicode=true,          
  plainpages=false,
  colorlinks=false,       
  citecolor=blue,        
}
\usepackage{orcidlink}
\usepackage{aliascnt}
\newaliascnt{eqfloat}{equation}
\newfloat{eqfloat}{h}{eqflts}
\floatname{eqfloat}{Equation}

\newcommand*{\ORGeqfloat}{}
\let\ORGeqfloat\eqfloat
\def\eqfloat{%
  \let\ORIGINALcaption\caption
  \def\caption{%
    \addtocounter{equation}{-1}%
    \ORIGINALcaption
  }%
  \ORGeqfloat
}

\usepackage{cleveref}

\graphicspath{{Figures/}}

\author{Samuel J. Pitman \orcidlink{0009-0004-0467-9695}}
\author{Alicia K. Evans}
\author{Robbie T. Ireland \orcidlink{0000-0002-4282-7810}}
\author{Felix Lempriere}
\author{Laura K. McKemmish \orcidlink{0000-0003-1039-2143}}
\email{l.mckemmish@unsw.edu.au}
\affiliation[University of New South Wales]
{School of Chemistry, University of New South Wales, Sydney, NSW, 2052, Australia}

\title{Benchmarking Basis Sets for Density Functional Theory Thermochemistry Calculations: Why unpolarised basis sets and the  polarised 6-311G family should be avoided.}

\begin{document}


\begin{abstract}
Basis sets are a crucial but often largely overlooked choice when setting up quantum chemistry calculations. The choice of basis set can be critical in determining the accuracy and calculation time of your quantum chemistry calculations. Clear recommendations based on thorough benchmarking are essential, but not readily available currently. 

This study investigates the relative quality of basis sets for general properties by benchmarking basis set performance for a diverse set of 136 reactions (from the diet-150-GMTKN55 dataset). In our analysis, we find the distributions of errors are often significantly non-Gaussian, meaning that the joint consideration of median errors, mean absolute errors and outlier statistics is helpful to provide a holistic understanding of basis set performance. 

Our direct comparison of performance between most modern basis sets provides quantitative evidence for basis set recommendations that broadly align with the established understanding of basis set experts and is evident in the design of modern basis sets. For example, while zeta is a good measure of quality, it is not the only determining factor for an accurate calculation with unpolarised double and triple-zeta basis sets (like 6-31G and 6-311G) having very poor performance. Appropriate use of polarisation functions (e.g. 6-31G*) is essential to obtain the accuracy offered by double or triple zeta basis sets. 

In our study, the best performance in our study for double and triple zeta basis set are 6-31++G** and pcseg-2 respectively. However, the performance of singly-polarised double-zeta and doubly-polarised triple-zeta basis sets are quite similar with one key exception: the polarised 6-311G basis set family has poor parameterisation which means its performance is more like a double-zeta than triple-zeta basis set. All versions of the 6-311G basis set family should be avoided entirely for valence chemistry calculations moving forward. 


\end{abstract}

\singlespacing

\section{Introduction}

The uptake of computational quantum chemistry by new users to support discovery and innovation throughout chemistry and related sciences has never been higher. To ensure these new users maximise the effectiveness of their computational time and produce the highest quality models within their available resources for the systems they are investigating, users need robust recommendations on the technical specifications they should use for their calculations and advice on the expected errors of these calculations. 

Computational quantum chemistry is the field where electron distributions in molecules are explicitly modelled by solving the Schr\"odinger equation to varying levels of approximation\cite{SzaboText,jensen_introduction_2017}. Fortunately for users, the various approximations have been exceptionally well categorised and for gas-phase systems can typically be divided into (1) a level of theory or method (for most users a density functional approximation) that simplifies the equation into something solvable, and (2) a basis set that is used to mathematically represent the distribution of electrons in the system of interest. Together, the choice of level of theory and basis set are known as the model chemistry.  

The selection of model chemistry is crucial for determining the accuracy of the computational prediction as well as the calculation time\cite{atkins_physical_2014}. Many extensive studies\cite{goerigk_thorough_2011,goerigk_look_2017,2017HeadGordon,goerigk_trip_2019} exist evaluating the performance of different density functional approximations against high level benchmark results, providing quantitative general evidence for users to guide their selection of method. However, similar benchmarking studies for basis set performance are largely absent from the literature.

Experts on basis set design can typically advise on the strengths and limitations of particular basis sets; for example, that anions need diffuse functions\cite{clark_efficient_1983, Davidson_Basis_1986, JensenPolarisationIII2002b}, the 6-311G family are not valence-triple-zeta basis sets\cite{grev_6-311g_1989}, and property-specialised basis sets \cite{IrelandSpecialisation2023} should be used for properties which have unusual basis set demands such as NMR parameters\cite{KutzelniggIGLO1990,PeraltaDensity2004,jensen_basis_2006,DengCalculation2006,Jensen2008,Benedikt_Optimization_2008,ProvasiSeries2010,KjrPople2011,jensen_segmented_2015,aggelund_development_2018}, EPR parameters \cite{RegaDevelopment1996,BaroneRecent2002,BaroneDevelopment2008,BaroneValidation2009,HedegardOptimized2011,jakobsen_probing_2019} and X-ray \cite{Hanson-HeineBasis2018,FoudaAssesment2018,ambroise2018probing,FoersterTime2020,Ambroise_Probing_2021} calculations to name but a few. Such recommendations are based on a detailed understanding basis sets design and being able to read and understand the composition of basis sets in terms of primitive exponents and contraction schemes.

Current recommendations for the choice of basis set are largely theoretically motivated, with often limited quantitative evidence. The goal of this benchmark study is thus to provide quantitative evidence for basis set selection by analysing the performance of multiple basis sets for general-purpose chemistry reaction energies. Our focus in this benchmark is on providing comprehensive coverage of basis set performance for a diverse set of reactions. However, we consider only a relatively small number of reactions (139) and three diverse and representative hybrid density functional approximations. 

Our intent is that this study provides the preliminary data and results to enable smart design of a more extensive and targeted follow-up benchmark study. In particular, the purpose of this  study is to:
\begin{itemize}
    \item identify the most important considerations when selecting basis set and determining basis set errors; 
    \item identify aspects of basis set choice that can, for most valence chemistry calculations, be ignored;
    \item make a preliminary assessment of the relative importance of the choice of basis set and the choice of hybrid density functional approximation;
    \item make a preliminary assessment of whether the popular density functional theory (DFT) benchmarking rankings, done at the complete basis set limit, are reliable indications of performance for more modest basis sets.  
\end{itemize} 

Section 2 is a brief introduction to important basis set terminology, and basis set design which may be skipped by a familiar reader. Section 3 describes the methodology used within this benchmark including the selection of a benchmark data set, model chemistry's, calculation details and the statistics used to compare basis set performance. Section 4 discusses the results of the benchmark, comparing unpolarised and polarised basis sets, the benefits of diffuse augmentation and timing considerations. Section 5 concludes and provides future directions of work.

\section{Basis Sets in Quantum Chemistry}
It is worth introducing important basis set terminology and briefly reviewing what is currently known about basis set design, referring the reader to recent reviews\cite{jensen2013atomic,hill2013gaussian,nagy2017basis} for detailed discussions.

A basis set is a predetermined set of atomic basis functions, $\chi(r)$, with different angular momentum (e.g. $s$, $p$, $d$) and different composition. Molecular orbitals are constructed by linear combination of atomic basis functions, i.e.
\begin{equation}
\phi_i(r) = \sum_j c_{ij} \chi_j(r) \ ,
\end{equation} where $c_{ij}$ are the coefficients optimised usually by a self-consistent-field (Hartree-Fock or DFT) calculation for a specific molecule.

Whilst many kinds of basis sets exist \cite{slater_atomic_1930, JaffardWavelets2001, HarrisonMultiresolution2003, HarrisonMultiresolution2004, KittelIntroduction2005}, by far the most popular basis sets used today are those based on primitive Gaussian basis functions \cite{boys_electronic_1950}, which have a radial function of the form
\begin{equation} 
    G_\alpha(r) = N e^{-\alpha r^2} \ ,
\end{equation} where $N$ is the normalisation constant, $\alpha$ is known as the Gaussian exponent and $r$ represents the radial separation between particles. The angular momentum component of the basis set is defined through the appropriate spherical harmonic. 

Gaussian functions, despite failing to describe the electron-nuclear cusp \cite{kato_eigenfunctions_1957} and long-range behaviour \cite{mayer_simple_2003} are ubiquitous with quantum chemistry on account of their computational efficiency. Computational costs can be further reduced by contracting primitive Gaussian function basis, whereby atomic basis basis functions can be written in terms of a linear combination of Gaussian primitives. The radial components of atomic basis functions can be defined as 
\begin{equation} 
\chi_j(r) = \sum_k d_{jk}  G_{\alpha_k}(r) \ .
\end{equation} 
The coefficients $d_{jk}$ are known as the contraction coefficients, with $k$ being the number of primitives within the contracted basis function. When each Gaussian primitive only contributes to a single Gaussian basis function, the basis set is said to have segmented contraction. Otherwise, all primitives contribute varying amounts to all contracted functions and the basis set is generally contracted\cite{jensen_unifying_2014}. It should be noted that most basis sets are neither fully segmented or generally contracted, instead opting to increase the flexibility of the functions representing the valence shell by leaving them uncontracted. This results in a contraction scheme such as (10$s$,6$p$,2$d$,1$f$) $\to$ [4$s$,3$p$,2$d$,1$f$] (for carbon pc-2) where the final (i.e. lowest exponent) two $s$ functions, the final two $p$, the two $d$ and single $f$ functions remain uncontracted. The relative speed of segmented and generally-contracted basis sets depends on how the integral evaluation package is implemented within a particular quantum chemistry package; most programs have been written with segmented basis sets in mind \cite{jensen2013atomic}. 

Basis sets are defined for a set of individual elements, based on designing a basis set structure (i.e. number and type of primitive and contracted functions) and optimising function parameters (e.g. in variational atomic or molecular calculations) in order to produce a useful set of building blocks for describing electron distributions in a variety of chemical systems. Different basis sets have been optimised in different ways and for different purposes, e.g. the correlation-consistent basis sets (e.g. cc-pV$n$Z) \cite{dunning_gaussianI_1970,dunning_gaussianI_1989,KendallElectron1992,WoonCalculation1993,WoonGaussianIV1994,WoonGaussianV1995,WilsonGaussianVI1996,PrascherGaussisanVII2011,VanMourikGaussianVIII2000,WilsonGaussianIX1999,DunningGaussianX2001} were optimised for correlation energies while the polarisation consistent pc(seg)-$n$ \cite{jensen_polarization_2001, JensenPolarizationII2002a, JensenPolarisationIII2002b,JensenPolarizationIV2003, jensen_polarizationV_2004, jensen_polarization_2007} basis sets were optimised for HF and DFT energies. 

Individual basis functions are mathematical functions that together are used to construct the molecular orbitals, but it can be useful to label individual basis functions in terms of the particular molecular orbital they contribute to most strongly. For example, most general-purpose all-Gaussian basis sets have a single highly contracted Gaussian function that is used primarily to describe the $1s$ core electrons; this is thus sometimes called the `core' basis function in the basis set. (e.g. "6" in 6-31G). 

To describe many properties in chemistry, including reaction energies and geometries, we need to mathematically describe the valence region around atoms in a sufficiently-flexible manner in order to accommodate orbital changes as different bonding occurs. For useful accuracy, this requirement typically means that instead of a single atomic basis function being the major contributor to each valence orbital, multiple basis functions are used in this way. A well-known example is the 6-31G basis, where each valence orbital is described by two valence functions; the first is written in terms of three primitive functions with the second is written terms of a single primitive function. This reasoning leads to the definition of `zeta', the number of basis functions used to represent each valence orbital within a basis set. For example, the 6-31G basis set is a double-zeta basis set as it uses two basis functions to describe each valence orbital. Since zeta often refers to the splitting of the valence orbitals, `valence-zeta' may be a more precise term here. `Zeta' is probably the most commonly used metric for basis set size and is typically used as a convenient shorthand for the quality and computational cost of a calculation. For many applications, double- or triple-zeta basis sets with hybrid density functionals are the standard approach, such as the popular B3LYP/6-31G* model chemistry.

As previously alluded to, changes in electron distribution occur as atoms bond to form molecules. Experience shows that there needs to be additional flexibility in the basis set description for atoms in order to provide orientation flexibility. In basis sets, this need is typically addressed by adding polarisation functions, which are basis functions of higher angular momentum than the occupied atomic orbitals for a particular element; a $p$ function is a polarisation function for hydrogen, whilst a $d$ function is a polarisation function for carbon. Modern basis set understanding teaches that polarisation functions should always be used when running calculations, i.e. that 6-31G should never be used and instead users should use 6-31G* (equivalently 6-31G(d)) or 6-31G** (6-31G(d,p)). The degree of polarisation is suggested as one less than the zeta-quality; for a double-zeta basis set, a single polarisation function should be added, while for triple-zeta, two levels of polarisation should be used. That is, unpolarised basis sets for the carbon atom have simply $s$ and $p$ functions; a singly-polarised basis set has additional $d$ functions only, whilst a doubly-polarised basis set has sets of additional $d$ functions and additional $f$ functions.

The inclusion of polarisation functions within the basis sets are now known to be critical for accurate calculations \cite{jensen_polarization_2001}, and all modern basis set families include these as an essential part of the basis set. The importance of polarisation functions is perhaps acknowledged most clearly in the naming convention of the polarisation consistent pc-$n$ basis set family, where $n$ is the level of polarisation not the zeta level. However, it is important for modern basis set users to know that early basis sets, most importantly the Pople basis set family, didn't follow modern approaches and instead treated polarisation functions as an add-on to the basis set, indicated by asterisks or parentheses (e.g. (d), (2df,p). 

When the electron cloud of a system is particularly large - such as in anionic systems - additional low-exponent Gaussian functions called diffuse functions need to be added to the basis set. The resulting basis sets are typically called diffuse augmented basis sets \cite{clark_efficient_1983} and are often required for the description of negatively charged species \cite{JensenPolarisationIII2002b}.
For species or atoms that do not meet these criteria, the impact of diffuse augmentation is typically inconsequential, though it has been noted that in some cases unnecessary addition of diffuse functions can lead to convergence problems \cite{Chandrasekhar_Efficient_1981, Papajak_Perspectives_2011, Bauza_Is_2013}. 

Basis set requirements for density functional approximations and wavefunctions methods are somewhat different, but details are beyond the scope of this introduction and paper. We note that hybrid density functional approximations include a Hartree-Fock (HF) component while double-hybrid density functional approximations include both a HF and MP2 component, meaning that the theoretical basis set requirements become complex. Here, we focus on evaluating basis set performance for hybrid and double-hybrid density functional approximations used with small to medium-size basis sets (i.e. single to triple-zeta basis sets). Note that higher levels of theory (such as variants of coupled-cluster theory) should usually use basis set extrapolation techniques \cite{varandas2018straightening}, and consider explicitly correlated basis sets \cite{ten2012explicitly,kong2012explicitly} and composite methods \cite{Karton2023} to ensure very high accuracy. 

This paper's benchmarking results are designed to support calculations on large molecules or large numbers of molecules (e.g. computationally generating data for machine learning approaches \cite{yang2019analyzing,keith2021combining}), for which computational resources do not permit these types of very high accuracy methodologies, and thus the choice of basis set is important. 

It is worth briefly describing the families of basis sets that are considered in this study.
\begin{itemize}
    \item \textbf{Pople:} The Pople family  is composed of the STO-$n$G \cite{Hehre1969, HehreSelf1970} and the $k$-$nlm$G \cite{Ditchfield1971,HehreExtended1972, HehreFurther1972} basis sets e.g. STO-3G and 6-31G. STO-$n$G are single zeta basis sets created by fitting $n$ Gaussians to the Slater-type orbital of an atom. The $k$-$nlm$G basis sets are known as `split-valence' basis sets, where $k$ is the number of Gaussians describing each core orbital, and $m$, $l$, $n$ designate the number of functions in the valence shell and so the number of `splits' in the valence shell (also known as zeta). Notably, the split-valence Pople basis sets have equal exponents for the $s$ and $p$ valence shell.
    \item \textbf{Correlation Consistent:} The correlation consistent (cc) basis sets \cite{dunning_gaussianI_1989}, also known as the Dunning family, are designed to recover the correlation energy of the valence electrons. These basis sets add sets of primitive functions based on their contribution to the correlation energy. For example, the first $d$ function contributes significantly more to the correlation energy, while the second $d$ function and first $f$ function contribute similarly. Thus the fist set of additional functions would be a single $d$ function, whilst the second set would add another $d$ function and a single $f$ function. 
    \item \textbf{Karlsruhe:} A commonly used family also known as the Ahlrichs basis sets, the def2 basis sets \cite{weigend_balanced_2005} are HF optimised with polarisation functions taken from the correlation consistent basis sets. Several examples of these basis sets are the def2-SVP, def2-TZVPD and def2-TZVPP. Def2-SVP is a double zeta basis set, with polarisation functions on hydrogen. The 'D' in def2-TZVPD indicates diffuse augmentation of the basis set while the additional 'P' in def2-TZVPP indicates an additional level of polarisation on hydrogen. 
    \item \textbf{Polarisation Consistent:} The polarisation consistent (pc-$n$) basis sets \cite{jensen_polarization_2001}, also known as the Jensen family, are DFT optimised basis sets. The pc family is developed analogously to the cc basis sets, however, they are optimised towards DFT energies. The level of polarisation is indicated by $n$, where $n = 0$ repersents no polarisation functions and further integers indicate an additional level of polarisation. 
\end{itemize}

\section{Methodology}

\subsection{Selection of Benchmark Data Set}
A good benchmark data set for evaluating general-purpose valence chemistry should be diverse, large and representative of molecules that would be of interest to practical users. In the literature, there are several high-quality benchmark sets, most notably  GMTKN55 \cite{goerigk_look_2017} and MGCDB84 \cite{2017HeadGordon}, although all were developed to method benchmarking. GMTKN55 was chosen due to its diverse set of thermochemistry, kinetics and noncovalent interaction reactions. Additionally, the diet-GMTKN55 benchmarking sets\cite{gould_diet_2018} also allow for a representative set of reactions to be calculated. This greatly reduces the computational cost of performing a benchmark producing similar results to GMTKN55. 

As this is the first extensive benchmarking study of basis set performance to our knowledge, our goal was to perform a fast, initial analysis across a very wide selection of basis sets to determine the most important broad trends, paving the way for future investigations to investigate subtle differences in basis set performance using a more extensive benchmarking data set. Thus, we started with a smaller version of the GMTKN55 benchmarking set, known as the diet-150-benchmarking set\cite{gould_diet_2018}, which was selected to have similar results to the full GMTKN55 benchmarking set.

Diet-150-GMTKN55 is composed of the same types of reactions as GMTKN55, being small and large system reaction energies, isomerisation reactions, reaction barrier heights, intermolecular noncovalent reactions and intramolecular noncovalent reactions. We removed 11 reactions to produce a new diet-139-GMTKN55 data set for the following reasons: 
\begin{description}
    \item[Many basis sets are undefined for unusual elements:] To maximise the comprehensiveness of basis set coverage, we removed reactions with elements bromine, bismuth, tellurium and antimony. 
    \item[Timing considerations:] In initial investigations, calculations for the \ce{C60} molecule dominated computation time but did not significant influence results and so the reaction involving \ce{C60} was removed. 
\end{description}
The diet-139-GMTKN55 benchmarking set contains a small number of negatively charged species (19), which are expected to be particularly sensitive to the inclusion of diffuse functions. 

The molecular geometries and reaction energy reference values were taken from the GMKTN55 database \cite{goerigk_look_2017}. The GMTKN55 reference values have been carefully considered for each of its component datasets, with details provided in Table 1 of \citet{goerigk_look_2017}. Overall, the reference values are usually very high-level quantum chemistry methods like CCSD(T)/CBS, but there are also some theoretically back-corrected experimental values.

\begin{table}[htbp!]
\centering
\footnotesize
\caption{Basis sets benchmarked in this study. All timings for the full HF calculation (i.e. not a single SCF cycle) relative to the 6-31G** basis set (bold) and should be regarded as indicative only. The ``Singular'' and ``Doubly'' polarisation terms denote one and two levels of polarisation, respectively, for non-hydrogen species, while ``H'' and ``2H'' denote one or two levels of polarisation for hydrogen. Contraction schemes are provided as well as whether diffuse augmentation functions have been added. }
\begin{tabular}{c c H c c c c c c} 
 \toprule
\textbf{Basis Set} & \textbf{Reference} & \textbf{Family} &\textbf{Zeta} & \textbf{Polarisation} & \textbf{Primitives} & \textbf{Contraction} & \textbf{Diffuse} & \textbf{Rel Timing} \\ [0.5ex] 

\vspace{-0.5em} \\
\textit{\underline{Very fast}} \\
 STO-3G & Refs. \citenum{Hehre1969}, \citenum{HehreSelf1970} & Pople & Single & None & 6s3p & 2s1p & No & 0.45\\
 3-21G & Ref. \citenum{BinkleySelf1980} & Pople & Double & None & 6s3p & 3s2p & No & 0.55\\
 pc-0 & Ref. \citenum{jensen_polarization_2001} & Jensen & Double & None & 5s3p & 3s2p & No & 0.56\\
 pcseg-0 & Ref. \citenum{jensen_unifying_2014} & Jensen & Double & None & 6s3p & 3s2p & No & 0.59\\
 6-31G & Ref. \citenum{HehreFurther1972} & Pople & Double & None & 10s4p & 4s3p & No & 0.63\\

 6-311G & Ref. \citenum{Krishnan1980SelfFunctions} & Pople & Triple & None & 11s5p & 4s3p & No & 0.84\\
 6-31+G & Ref. \citenum{Frisch1984SelfSets} & Pople & Double & None & 11s5p & 4s3p & Yes & 0.88\\
 aug-pc-0 & Ref. \citenum{JensenPolarisationIII2002b} & Jensen & Double & None & 6s4p & 4s3p & Yes & 0.95\\
 6-31++G & Ref.  \citenum{Frisch1984SelfSets} & Pople & Double & None & 11s5p & 4s3p & Yes & 0.96 \\
 aug-pcseg-0 & Ref. \citenum{jensen_unifying_2014} & Jensen & Double & None & 7s4p & 4s3p & Yes & 0.99\\
\vspace{-0.5em} \\
\textit{\underline{Fast}} \\
\textbf{6-31G*} &\textbf{Ref. \citenum{HariharanInfluence1973}} & \textbf{Pople} &\textbf{Double} & \textbf{Singular} & \textbf{10s4p1d} & \textbf{3s2p1d} & \textbf{No} & \textbf{1.0}\\
 6-31G** & Ref. \citenum{HariharanInfluence1973}  & Pople & Double & Singular, H & 10s4p1d & 3s2p1d & No & 1.2\\
 def2-SVP & Ref. \citenum{weigend_balanced_2005} & Karlsruhe & Double & Singular, H & 7s4p1d & 3s2p1d & No & 1.2\\
 6-311+G & Ref. \citenum{Frisch1984SelfSets} & Pople & Triple & None & 12s6p & 5s4p & Yes & 1.2\\
 pcseg-1 & Ref. \citenum{jensen_unifying_2014} & Jensen & Double & Singular, H & 8s4p1d & 3s2p1d & No & 1.3 \\
 6-311G* & Ref. \citenum{Krishnan1980SelfFunctions} & Pople & Triple & Singular & 11s5p1d & 4s3p1d & No & 1.3\\
 pc-1 & Ref. \citenum{jensen_polarization_2001} & Jensen & Double & Singular, H & 7s4p1d & 3s2p1d & No & 1.3 \\
  6-31+G* & Ref. \citenum{Frisch1984SelfSets} & Pople & Double & Singular & 11s5p1d & 4s3p1d & Yes & 1.4\\

\vspace{-0.5em} \\
\textit{\underline{Moderate}} \\
 cc-pVDZ & Ref. \citenum{dunning_gaussianI_1989} & Dunning & Double & Singular, H & 9s4p1d & 3s2p1d & No & 1.5\\
 6-311G** & Ref. \citenum{Krishnan1980SelfFunctions} & Pople & Triple & Singular, H & 11s5p1d & 4s3p1d & No & 1.5 \\

 6-31++G** & Ref. \citenum{Frisch1984SelfSets} & Pople & Double & Singular, H & 11s5p1d & 4s3p1d & Yes & 1.8\\
 6-311+G* & Ref. \citenum{Frisch1984SelfSets} & Pople & Triple & Singular & 12s6p1d & 5s4p1d & Yes & 1.9\\
 6-31G(2df,p) & Ref. \citenum{Frisch1984SelfSets} & Pople & Double & Doubly, H & 10s4p2d1f & 3s2p2d1f & No & 3.3\\
 def2-SVPD & Ref. \citenum{Furche2010Property} & Karlsruhe & Double & Singular, H & 8s4p2d & 4s2p2d & Yes & 3.5\\
 6-311G(2df,p) & Ref. \citenum{Frisch1984SelfSets} & Pople & Triple & Doubly, H & 11s5p2d1f & 4s3p2d1f & No & 4.3\\

\vspace{-0.5em} \\
\textit{\underline{Slow}} \\
 aug-pcseg-1 & Ref. \citenum{jensen_unifying_2014} & Jensen & Double & Singular, H & 9s5p2d & 4s3p2d & Yes & 5.3\\
 aug-pc-1 & Ref. \citenum{JensenPolarisationIII2002b} & Jensen & Double & Singular, H & 8s5p2d & 4s3p2d & Yes & 5.7\\
 aug-cc-pVDZ & Ref. \citenum{KendallElectron1992} & Dunning & Double & Singular, H & 10s5p2d & 4s3p2d & Yes & 6.4\\
 def2-TZVP & Ref. \citenum{weigend_balanced_2005} & Karlsruhe & Triple & Doubly, H & 11s6p2d1f & 5s3p2d1f & No & 6.2\\
 pcseg-2 & Ref. \citenum{jensen_unifying_2014} & Jensen & Triple & Doubly, 2H & 11s6p2d1f & 4s3p2d1f & No & 9.2\\
 cc-pVTZ & Ref. \citenum{dunning_gaussianI_1989} & Dunning & Triple & Doubly, 2H & 10s5p2d1f & 4s3p2d1f & No & 9.4\\
 def2-TZVPP & Ref. \citenum{weigend_balanced_2005} & Karlsruhe & Triple & Doubly, 2H & 11s6p2d1f & 5s3p2d1f & No & 9.6\\
 pc-2 & Ref. \citenum{jensen_polarization_2001} & Jensen & Triple & Doubly, 2H & 10s6p2d1f & 4s3p2d1f & No & 9.9\\
 \vspace{-0.5em} \\
\textit{\underline{Very Slow}} \\
 def2-TZVPD & Ref. \citenum{Furche2010Property} & Karlsruhe & Triple & Doubly, H & 12s6p3d1f & 6s3p3d1f & Yes & 37 (17)$^\dagger$\\
 aug-cc-pVTZ & Ref. \citenum{KendallElectron1992} & Dunning & Triple & Doubly, 2H & 11s6p3d2f & 5s4p3d2f & Yes & 138  (74)$^\dagger$\\
 aug-pc-2 & Ref. \citenum{JensenPolarisationIII2002b} & Jensen & Triple & Doubly, 2H & 11s7p3d2f & 5s4p3d2f & Yes & 153 (81) $^\dagger$\\
[1ex] 
 \bottomrule
\end{tabular}
\label{t:basissetsGB}

$^\dagger$ All timings were calculated with standard integral accuracy apart from those in parenthesise, where higher integral accuracy thresholds that led to convergence in fewer SCF cycles were used. \end{table}

\subsection{Model Chemistry Selection} 

We aimed for very broad coverage of popular basis sets for prediction of general-property valence chemistry from single- to triple-zeta quality and from unpolarised to doubly-polarised. The basis sets considered in this benchmarking study are detailed in Table \ref{t:basissetsGB}, where the zeta, polarisation and diffuse augmentation are specified for each basis set. The basis sets are grouped into five different speed tiers based on their approximate relative timings for a representative set of 25 molecules (see Sec. \ref{SecTimings} for more details). The Pople, correlation consistent, polarisation consistent (both generally and segmented contracted versions) and Karlsruhe basis sets were all represented. All basis sets were designed for general-purpose chemistry, meaning no basis sets specialised towards core-dependent properties (e.g. pcS-$n$, pcJ-$n$) were considered. Further, no quadruple- or quintuple- basis sets were considered as a  {T,Q} or {Q,5} complete basis set extrapolation with very small errors will yield predictions with accuracies well below the density functional approximation error.

This paper's goal is to focus on basis set performance, not functional performance. Therefore, these functionals are selected largely to be illustrative only and not to provide specific recommendations; we refer readers to more thorough density functional benchmark studies for detailed analysis of their performance\cite{Korth_Mindless_2009, goerigk_thorough_2011, 2017HeadGordon,goerigk_trip_2019}. 
The three DFT methods chosen were $\omega$B97M-V, M06-2X and B3LYP and are all hybrid level accuracy. The B3LYP\cite{becke_new_1993, lee_development_1988} functional was chosen as, despite not being the most actuate functional, it is commonly used \cite{santra_observations_2019}. On the other hand, the $\omega$B97M-V\cite{mardirossian_b97m-v_2016} and M06-2X\cite{zhao_improved_2008} are generally considered to be more accurate methods \cite{goerigk_thorough_2011, santra_observations_2019}. The decision to use three density functionals was made to ensure that the conclusions drawn can be confidently attributed towards the performance of the basis sets, rather than due to the performance of a particular functional.

\subsection{Calculation Details}\label{GPCalcDetails}

The single point energies were calculated using Q-chem 5.4.2 \cite{shao_advances_2015}  for diet-139-GMTKN55 using all basis sets given in Table \ref{t:basissetsGB} and the three different levels of density functional approximations. Geometries were taken from the GMKTN55 database.

For large diffuse augmented basis sets, there were some molecules for which convergence was very difficult, a known issue usually due to linear dependence issues resulting from diffuse functions being unnecessary to describe the molecule\cite{Chandrasekhar_Efficient_1981, Papajak_Perspectives_2011, Bauza_Is_2013}. 
In these cases, the result from the basis set lacking the diffuse augmentation was sometimes adopted. The molecules affected are: ISOL24 (i2p, i4e, i4p), UPU23 (1b, 2p, 1g) and MCONF (30); these are large molecules, specifically part of the following datasets - ISOL24 are isomerisation energies of large organic molecules, UPU23 is a set of relative energies between RNA-backbone conformers,  and MCONF are relative energies in melatonin conformers\cite{goerigk_look_2017}.  

\subsection{Statistical Analysis of Basis Set Performance}\label{StatsGB}

For each model chemistry, we obtained a set of 139 differences between the calculated result and the benchmark value which we refer to as errors for ease of reference.  We evaluate the statistical distribution of these errors in a number of ways for our results which we outline below.

Perhaps the most common statistic used to define basis set accuracy is the mean absolute error (MAE) or deviation (MAD), where absolute deviations from a benchmark value are found and then averaged for all calculations involving the basis set. However, we find our error distributions are often significantly skewed/ non-Gaussian and that these MAEs are reasonably sensitive to presence of outliers. Median absolute errors (medians hereafter) are less affected by outliers in the data set and as such can be seen to represent more typical behaviour. In our data set, medians are typically 2-3 times smaller than MAEs (for a true Gaussian distribution, the median and MAE are identical). Another common statistic is the root-mean-squared errors (RMSEs), though these are not considered in this work as they are more susceptible to outlier results than MAEs. 

In order to gain some insight into the prevalence of outliers, various additional measures can be employed. Namely, we quantify (1) the outlier count, which is the proportion of calculations with errors above 20 \kJmol (which are not useful for most practical purposes) and (2) a 95\% error cut-off, which provides a good estimate for the maximum expected error in most situations. 

After preliminary assessments to exclude clearly poorly performing basis sets, it is worthwhile to explicitly show the statistical distribution of errors. Many of the statistics described previously can be clearly visualised using box-and-whisker plots, where the box shows the 25\% and 75\% quartiles of errors, a central line shows the median and the whiskers denote the range of the 5\% and 95\% error cut-off.

Due to the number of statistics used in this work, it is useful to have a single number in order to provide a complete assessment of basis set quality. Similar combined statistics, whilst being found in other fields (e.g. Ref. \citenum{best2015plumbing}), has not yet been used in quantum chemistry benchmark studies to the best of our knowledge. In the context of our benchmark study, we define the performance statistic as a metric that defines basis set performance by considering the other statistical measures previously described. Specifically, the performance statistic averages the normalised performance of the MAE, median, outlier count and and  95\% error cut-off across the full set of basis sets considered. In turn, the performance statistic for a basis set is quoted as a single value indicative of its performance, where smaller values imply more strongly-performing basis sets.

\section{Results and Discussion}
\label{ResultsGB}

\subsection{Evaluating Basis Set Accuracy}

\begin{figure}[htbp!]
\centering
\includegraphics[width=0.95\textwidth]{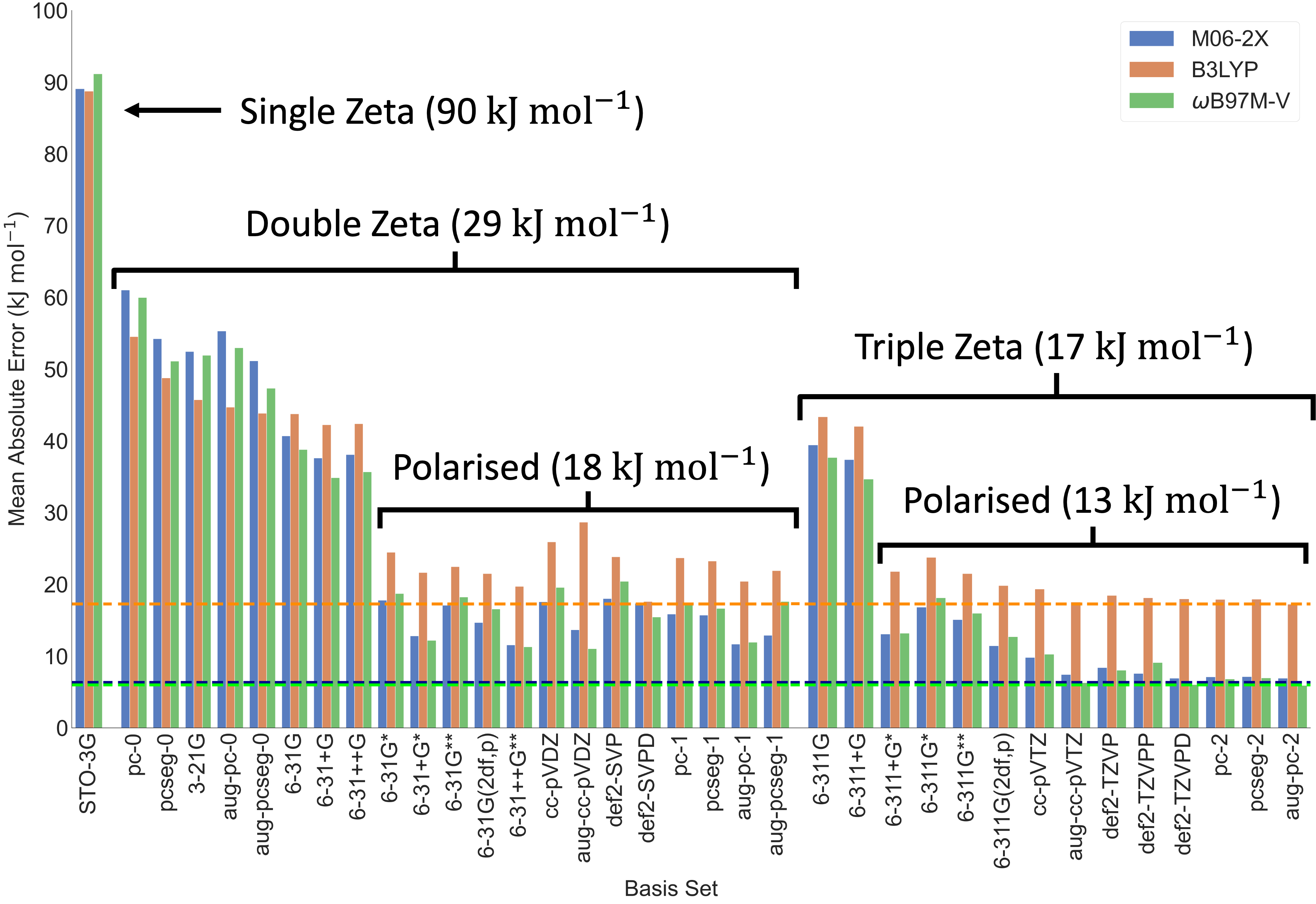}
\caption{A comparison of mean absolute errors between basis sets. 
The coloured horizontal dashed lines indicate the MAE complete basis set limit for each method, the best achievable value (ignoring error cancellations). The value in brackets is the average MAE for the stated category.}
\label{fig:general benchmark}
\vspace{1.5em}

\centering
\includegraphics[width=0.75\textwidth]{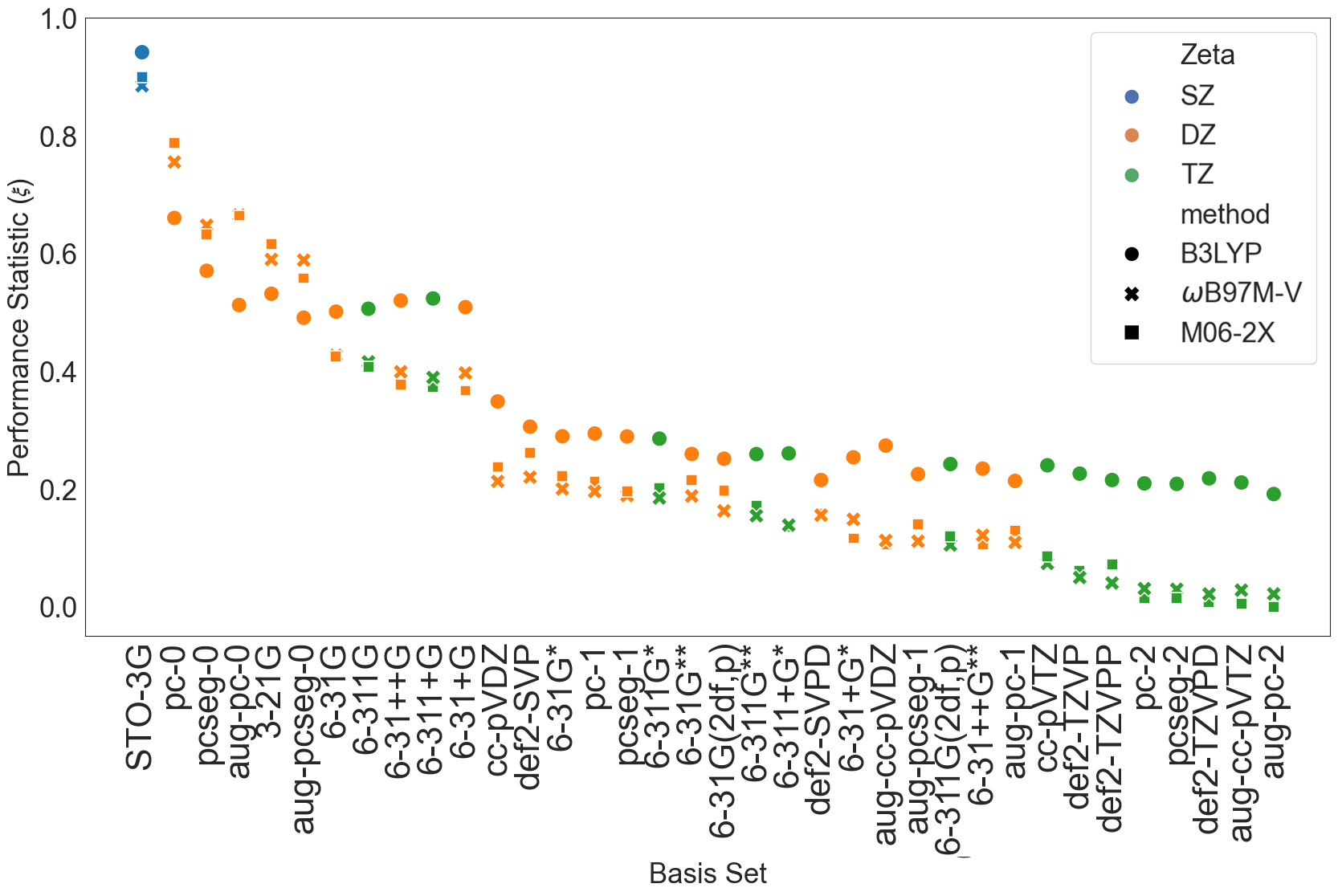}
\caption{The performance statistic of each basis set, with colours representing the zeta and shape indicating the method. A lower value indicates a model chemistry with more accurate calculations and fewer outliers.}
\label{fig:PerformanceStatistic}
\end{figure}

The overall benchmarking results for all basis sets under consideration are presented in \Cref{fig:general benchmark} (showing the average mean absolute errors across all 139 reactions for each basis set under consideration, visually separating by zeta-level and polarisation) and in \Cref{fig:PerformanceStatistic} (showing the performance statistic combining the MAE with median, outlier count and 95\% error cutoff to provide a more holistic understanding of the error in a single number). 

\begin{figure}[htpb!]
\centering
\includegraphics[width=1\textwidth]{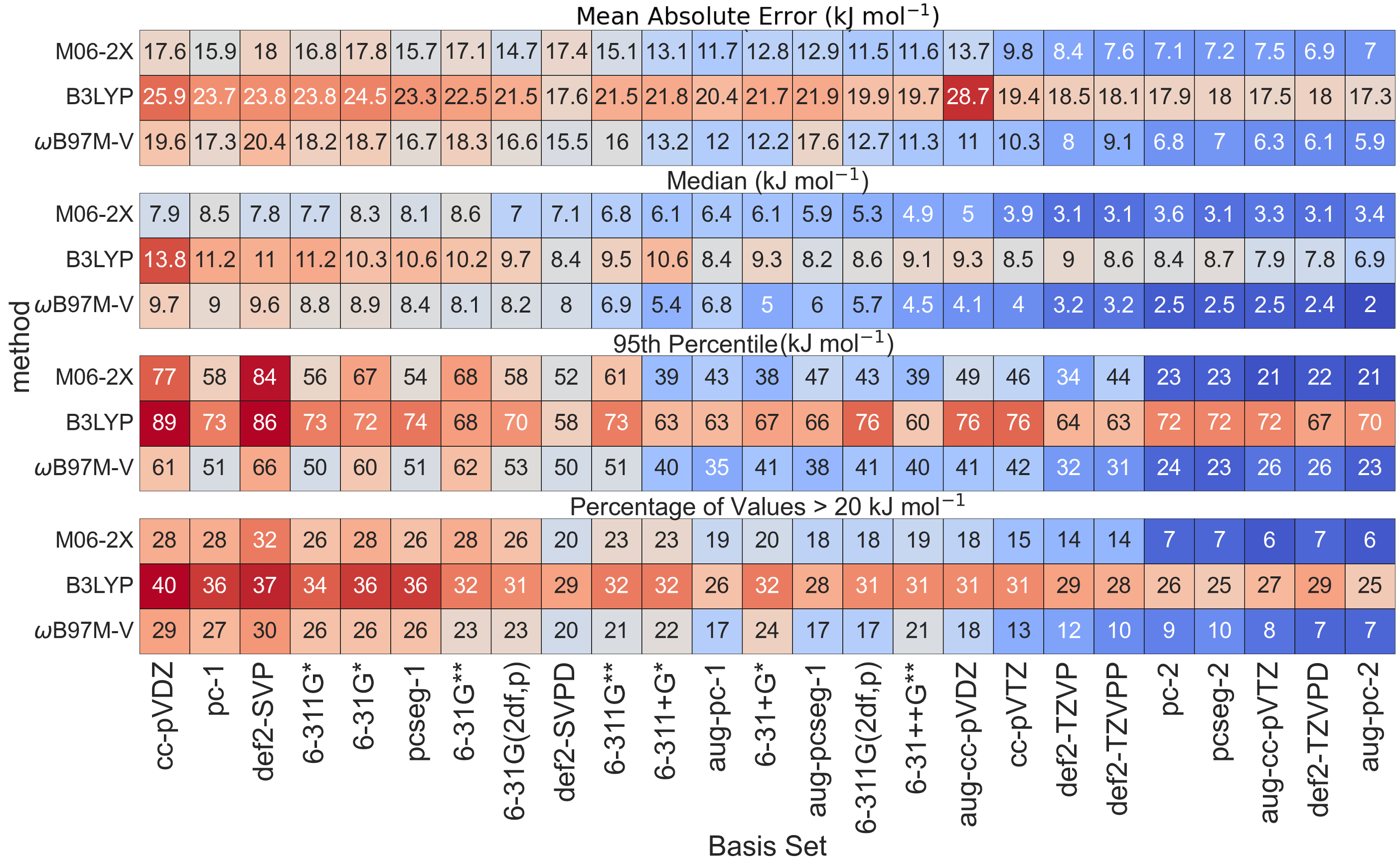}
\caption{A heat map of the mean, median, 95th percentile and the percentage of values above 20 \kJmol for absolute errors (MAEs) from the benchmark value for all polarised double and triple zeta polarised basis sets in \kJmol. Basis sets are ordered by an average of the MAEs across all methods. Red shading indicates very poor performance, and dark blue very good performance. The MAE complete basis set (CBS) limits are 6.4, 17.8, 6.2 \kJmol for M06-2X, B3LYP and $\mathrm{\omega}$B97M-V respectively. The median CBS limits are 2.8, 6.9, 2.0 \kJmol for M06-2X, B3LYP and $\mathrm{\omega}$B97M-V respectively.}
\label{fig:MeanandMedianHeatmaps}
\end{figure}

\begin{figure}[htpb!]
\centering
\includegraphics[width=0.85\textwidth]{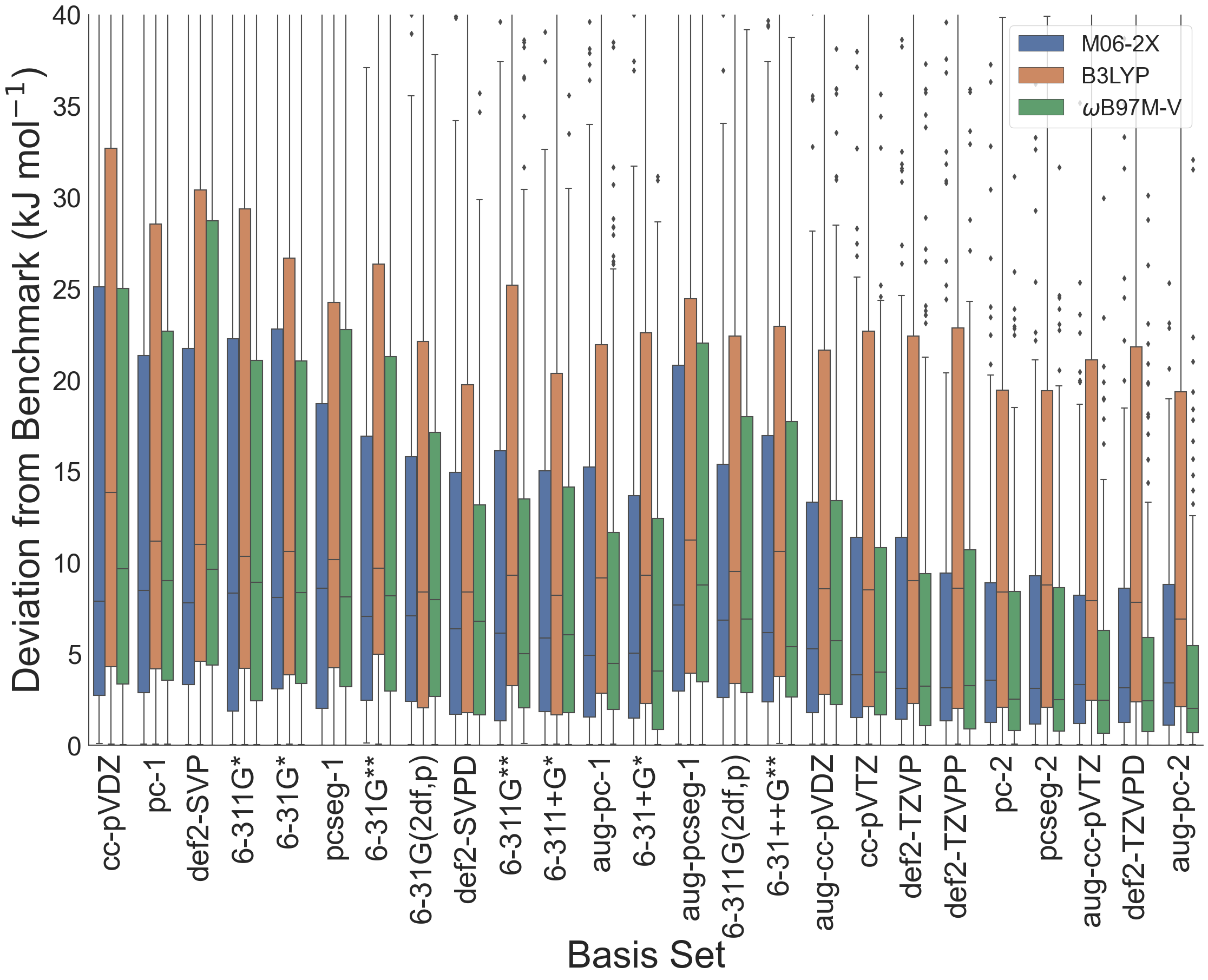}
\caption{A box and whisker plot comparing the performance of selected basis sets, with x axis ordering as in \Cref{fig:MeanandMedianHeatmaps}. }
\label{fig:DoubleZetaBoxPlot}
\end{figure}

For stronger performing basis sets or those of particular interest, we provide further details.

 Specifically, \Cref{fig:MeanandMedianHeatmaps}  quantifies the mean absolute errors, medians, 95\% cut-off and percentage of values with errors > 20 kJ/mol; this is important quantiative information about the likely accuracy (median) and reliability (cut-off and percentage of outliers) for each model chemistry.  This information together makes up the performance statistic, but the individual components can be useful for different applications and also demonstrate the importance of considering multiple metrics. Of note here is that mean absolute errors are around 2-3 times the value of the median error. 

Finally, \Cref{fig:DoubleZetaBoxPlot} uses box-and-whisker plots to show the statistical distribution of errors explicitly for our benchmark dataset.

Based on these results, we can make the following significant insights regarding the performance of the basis set (leaving functionals considerations for a later section). We utilise the $\omega$B97M-V medians as a convenient metric here, but we note that our key observations summarised here are usually consistent across different functionals and statistical metrics. 

In rough order of importance for our result dataset:
\begin{itemize}[listparindent=1.5em, labelsep=2em]

\item \textbf{Unpolarised} basis sets perform extremely poorly compared to their polarised counterparts, and there is little justification for their use in any application with hybrid DFT calculations. 
\item \textbf{With polarisation, zeta} is a reasonably reliable indication of basis set performance, i.e. polarised triple-zeta basis sets outperform polarised double-zeta basis sets in most cases.

\item \textbf{Polarised 6-311G Pople triple-zeta} basis sets exhibit significant under-performance compared to other triple-zeta basis sets (e.g. 6-311G(2df,p) at 5.6 \kJmol{} versus pc-2 at 2.5 \kJmol{} median error when employing the $\omega$B97M-V model chemistry). This consistent weak performance is observed across different polarisation choices. The subpar performance can be attributed to deficiencies in the optimization of $s$ contracted basis functions. In practice, the "3" basis function behaves more like a core basis function rather than a valence basis function, suggesting that 6-311G is better characterized as a double core-zeta, double valence-zeta basis set rather than a true triple valence-zeta. Although this basis set design may lead to unexpectedly strong performance in modelling core-dependent properties like NMR spectroscopy, it is recommended that property-specific  specialised basis sets be used for the calculation of core-dependent properties \cite{IrelandSpecialisation2023}.

\item \textbf{Diffuse functions} are very important to reducing errors for double-zeta basis sets in this data set (e.g. 6-31G* at 8.9 \kJmol{} vs 6-31+G* at 5 \kJmol{}). It is likely these errors are caused by a small number of large outliers in molecules (e.g. anions) where diffuse functions are critical for describing the larger electron distribution. Diffuse functions are relevant but less impactful for triple-zeta basis sets (e.g. pc-2 at 2.5 \kJmol{} vs aug-pc-2 at 2.0 \kJmol{}) as the additional valence functions can mimic the effect of the diffuse functions. \\ \hspace{1em} However, diffuse functions should only be used when the molecular system warrants its use as the timing increases are quite substantial (see below for discussion) and significant convergence issues can arise, especially for the diffuse augmented triple-zeta basis sets. Though these issues can be usually individually solved by increasing the accuracy of integral evaluation, ultimately this is additional computational time for no significant improvement in performance as the convergence issues arise because the additional basis functions are unnecessary to describe the electron distribution. \\ The question naturally arises, of course, on how users can determine when the molecular system warrants the use of diffuse functions. The prevalent wisdom is diffuse functions are needed for anions, but there are certainly counterexamples and other systems where diffuse functions are important. The underlying physical reason is that diffuse functions should be used when the electron distribution is significantly more extended (larger) than normal, for example, when electronegative atoms take on a strong partial negative charge, excited electronic states or simply when dissociation process is being considered modelled with a full potential energy surface. However, this approaches relies on chemical intuition; future work for automating decisions on when diffuse functions are and aren't necessary for a specific calculation would be very useful and could perhaps take advantage of the information provided by the linear dependency and convergence issues.     

\item \textbf{Polarised double-zeta} basis sets have variable errors similar in magnitude to the density functional approximation errors, with the inclusion or exclusion of diffuse functions being the most important factor determining performance. The relative performance of the different basis set families differed depending on the density functional and statistical metric considered, though we note aug-cc-pVDZ (4.1 \kJmol) and 6-311++G** (4.5 \kJmol) as high-performing double-zeta diffuse augmented basis sets while 6-31G** (8.1 \kJmol) and pcseg-1 (8.4 \kJmol) had good accuracy for double-zeta basis sets lacking diffuse functions. Timing considerations are thus likely to be an important consideration. 

\item \textbf{Non-Pople triple-zeta} basis sets all have errors very similar to the basis set limit results for a particular functional. Nevertheless, better performance is observed for pcseg-2 (2.5 \kJmol) compared to def2-TZVPP (3.2 \kJmol) or cc-pVTZ (4.0 \kJmol). This performance disparity can probably be attributed to the fact that pcseg-2 is optimized using a density functional (BLYP), whereas the other basis sets are optimized for wavefunction methods.

\item \textbf{Reduced polarisation functions on hydrogen} has a quite small impact on performance for our benchmarking set, especially for triple-zeta basis sets (for example, 6-31G* at 8.9 \kJmol{} vs 6-31G** at 8.1 \kJmol{} and def2-TZVP at 3.2 \kJmol{} vs def-TZVPP at 3.2 \kJmol{}). Appropriate combined use of pre-existing polarisation consistent and correlation consistent basis sets is likely to produce similar results (though further testing would be helpful here), e.g. calculations with pcseg-2 for non-hydrogens and pcseg-1 for hydrogens are likely to lead to faster calculations with minimal impact on performance. 

\end{itemize}

\subsection{Timing considerations}

\label{SecTimings}

When it comes to selecting a basis set, considering calculation times is crucial alongside accuracy considerations. However, obtaining precise timings can be surprisingly challenging due to the significant dependence on the molecular system and the specific computational chemistry package employed. For instance, in molecules without hydrogen atoms, the polarisation degree of the hydrogen becomes irrelevant to basis set timings, serving as a trivial example. Moreover, the timings can vary significantly between integral evaluation packages based on whether the package is optimized for general contraction (utilizing a primitive in multiple basis sets) or shared exponents between $s$ and $p$ basis functions. Additionally, the number of self-consistent field (SCF) cycles is algorithm-dependent and can be reduced with stronger initial guesses, such as those obtained from a smaller basis set. Nevertheless, it is important to note that the number of SCF cycles is not independent of the basis set; larger basis sets generally require more SCF cycles to achieve convergence.

Notwithstanding these caveats, we aim to provide insight into the relative calculation times of different basis sets. Specifically, Table \ref{t:basissetsGB} shows the relative timings for a full DFT calculation of a set of 25 representative molecular systems (from our basis set benchmarking list) compared to the 6-31G* basis set. The 25 molecules have an average number of total atoms of 15, where on average 7 are non-hydrogen atoms. Basis sets are categorised from "very fast" to "very slow", giving an indication of their calculation time.

As general trends for molecules of this size: 
\begin{itemize}
    \item \textbf{Unpolarised} basis sets were roughly half the computational time of their polarised equivalents;
    \item \textbf{Polarised} triple-zeta basis sets were roughly 1.5-5 times slower than polarised double-zeta basis sets, with the range largely determined by the number of additional polarisation functions;
     \item \textbf{Diffuse functions}, when included, usually increased computational cost significantly  especially for the polarisation consistent basis sets (roughly 4 times slower for pc-1 and 13 times slower for pc-2).  def2-TZVPD is significantly faster than the other diffuse augmented triple zeta basis set as it has one fewer $f$ basis function and a lower level of hydrogen polarisation. This timing increase can be attributed not only to the higher number of integrals but also to the increased SCF cycles resulting from the enhanced difficulty in SCF convergence with  diffuse functions: in fact, increasing the accuracy with which integrals were calculated actually decreased the calculation time because of the reduced number of SCF cycles. Thus the timings for diffuse functions will be very molecule-dependent, especially for the diffuse augmented triple-zeta basis sets with convergence specifics.   
\end{itemize}

\begin{figure}
\centering
\includegraphics[width=0.8\textwidth]{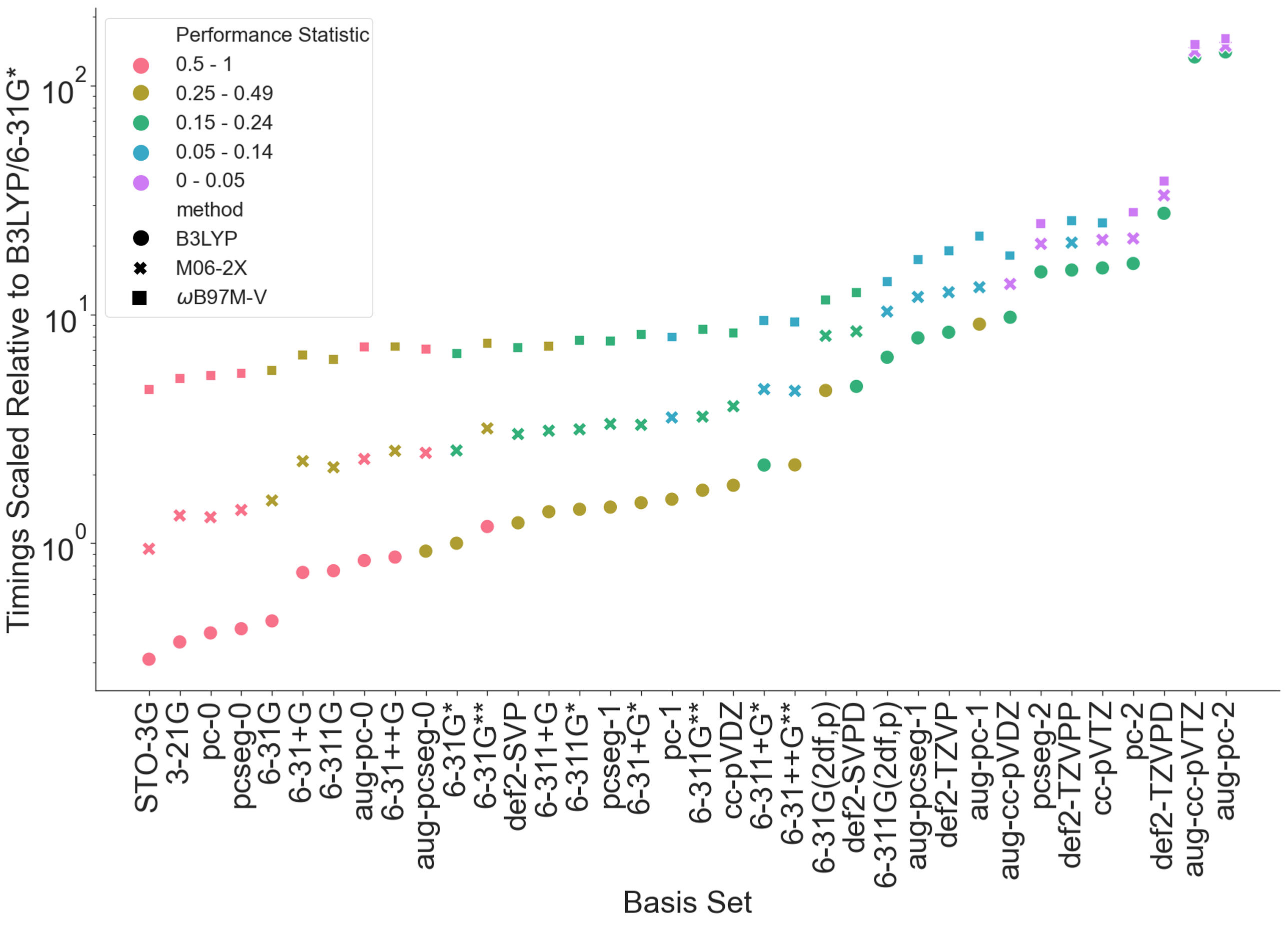}
\caption{The absolute timings of 25 molecules for each basis set with method indicated using a symbol. The y axis is the log of the mean absolute timings of the molecules. The performance statistic is included as a colour where a lower value indicates a better performance statistic. }
\label{fig:TimingsAndPStat}
\end{figure}

We combine timing and accuracy information visually in Figure \ref{fig:TimingsAndPStat}. 

The most immediate thing that stands out in these results is that the three density functional approximations differ considerably in their calculation time despite all being hybrid functionals on the same rung of Jacob's ladder: B3LYP is significantly faster than M06-2X which is again significantly faster than $\omega$B97M-V for our set of molecules. These results mean that for very large systems that are very demanding on computational resources, B3LYP calculations may remain feasible with double-zeta basis sets while the higher accuracy hybrids like M06-2X and $\omega$B97M-V aren't. This type of timing information, while known to experienced users and developers, is crucial for non-experts to understand yet rarely shared.



\subsection{Functional-dependence on basis set performance}

Our benchmark study is not intended to provide definitive functional recommendations; extensive benchmarking can be found, for example, within Refs.  \citenum{goerigk_thorough_2011,goerigk_look_2017} and \citenum{Goerigk2019}. Instead, we want to understand:
\begin{itemize}
    \item To what extent can complete-basis-set limit results from large density functional approximation benchmark studies be used to inform DFA performance for smaller basis sets? 
    \item Does functional choice influence which basis set performs most strongly? 
\end{itemize}

The findings presented in \Cref{fig:general benchmark} clearly demonstrate that benchmarking functionals near the complete basis set (CBS) limit cannot reliably predict functional performance for very small basis sets. In particular, it is evident that, though B3LYP has the poorest CBS performance, it outperforms the other two functionals when single-zeta and unpolarised double-zeta basis sets are used.

In contrast, \Cref{fig:general benchmark} clearly shows that the results obtained using non-Pople triple-zeta basis sets are bounded by the CBS limit, with basis set errors significantly less than density functional approximation errors. Therefore, the rankings in performance of hybrid density functional approximations obtained from a CBS limit benchmarking are likely to be applicable to non-Pople triple-zeta basis set calculations. 

When using polarised double-zeta basis sets, the magnitude of the functional error and basis set error are similar and it remains uncertain whether CBS limit performance alone can effectively guide the selection of the best functional. Given that the fastest reliable polarised triple-zeta basis set is nearly six times slower than its double-zeta counterpart, we recommend a future thorough benchmarking of DZ/hybrid DFT model chemistry's combinations. Such an approach is likely to yield valuable insights into identifying the most robust model chemistry methodologies. In this future study, it will also be important to take into account relative computational timings, as different hybrid functionals exhibit varying calculation times.

\subsection{Isolating basis set error}

The errors discussed thus far are dependent on both the density functional and the basis set employed. However, if the DFT/CBS limit is available, then we can separate these errors which can be helpful. Here, we do so for the M06-2X functional using the data from \citet{Goerigk2017}. 

Note that the basis set error is not independent of the functional, but this dependence is better considered in the total model chemistry error as has been done in the previous section. 

Furthermore, note that many discussions of basis set error often separate basis set incompleteness error (the error due to not using a `complete' (infinite) basis set) and basis set superposition error (the error created by atoms using the basis functions of nearby atoms to describe their own electron density). However, these errors can't be separated without further assumptions \cite{BSSEvsBSIE} which we will avoid in this discussion.

\begin{figure}
\centering
\includegraphics[width=1\textwidth]{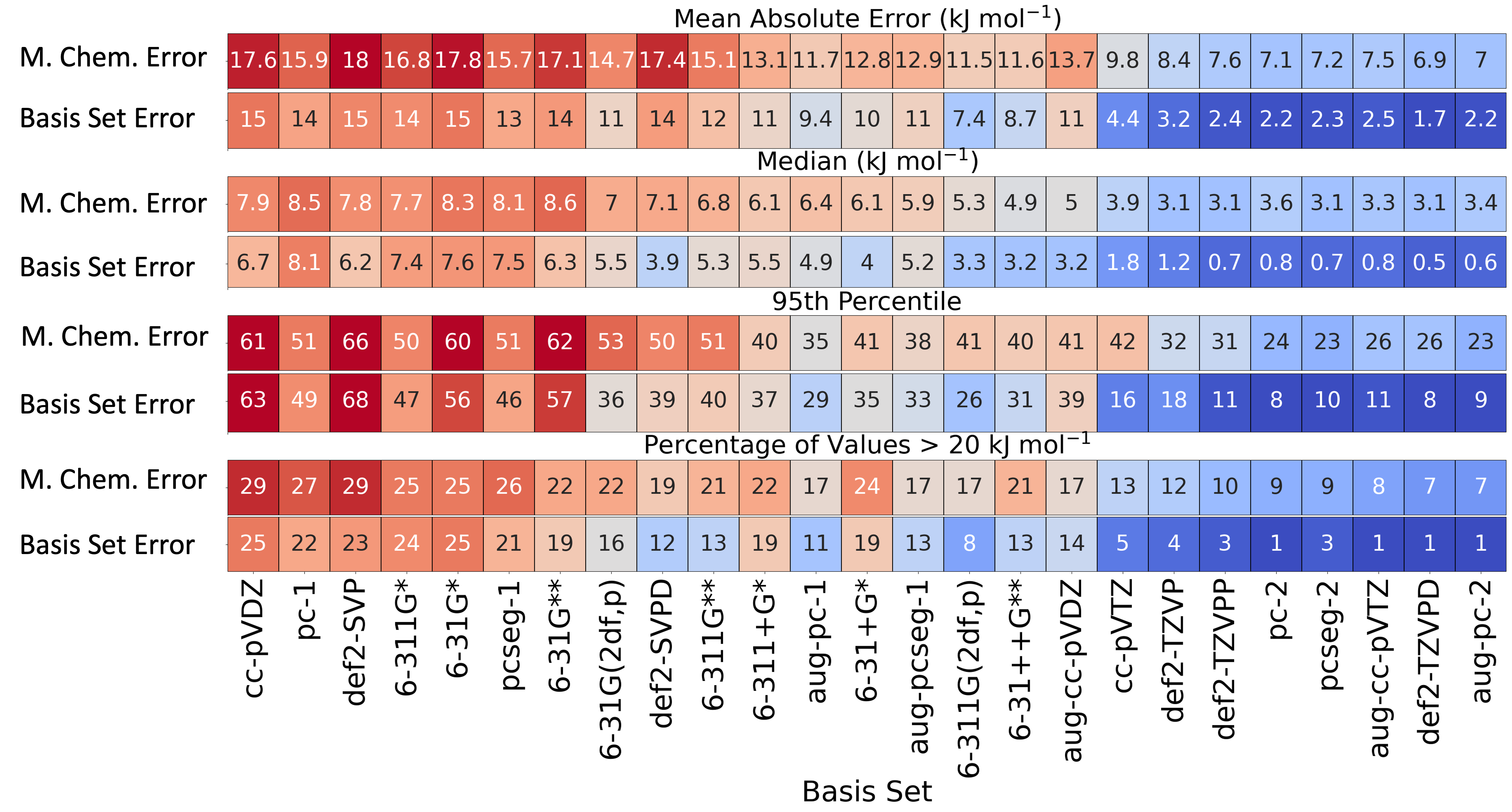}
\caption{A heat map of the model chemistry error and basis set error for M06-2X for all polarised double and triple zeta polarised basis sets, quantified by the mean, median, 95th percentile and the percentage of values above 20 \kJmol for absolute errors (MAEs) from the benchmark value  in \kJmol. For comparison, the M06-2X functional mean absolute error is 6.4 \kJmol and median absolute error is 2.8 \kJmol.  }
\label{fig:BSEmetrics}
\end{figure}


With these caveats, we present our results in Figure \ref{fig:BSEmetrics} that quantifies the model chemistry and basis set error for M06-2X using our four statistical metrics, reproducing the model chemistry error for ease of comparison. Note the functional error is constant across all basis sets, with a mean absolute error of 6.4 \kJmol{} and median absolute error of 2.8 \kJmol. Since we are working with statistical distributions of absolute errors, we have a triangular inequality, whereby the averaged absolute model chemistry errors are less than the sum of the averaged absolute density functional and basis set errors, with the difference being due to the cancellation of errors. 

For the best triple-zeta basis sets, the functional error is about a factor of three larger than the basis set error and so dominates, with the model chemistry error only slightly above the functional error. For double-zeta basis sets, the basis set error ranges from roughly equal to the functional error to twice as large, and therefore the model chemistry error is significantly higher than the functional error. For all model chemistries, we do see notable cancellation of errors between the functional and basis set errors, usually 2-6 \kJmol{} for mean absolute errors and 0-2 \kJmol{} for median absolute errors.

\section{Conclusions and Future Directions}

Here, we perform a comprehensive benchmarking of basis set performance for general-purpose ground state thermochemistry using 139 reactions of the diet-150-GMTKN55 data set, using three hybrid density functional approximations (M06-2X, B3LYP and $\omega$B97M-V) chosen for their diversity as representative. 

Our high-level recommendations are as follows:
\begin{itemize}
\item \textbf{Double zeta, singly polarised basis sets}: pcseg-1, or 6-31G** (with the $\omega$B97M-V method performing best in our limited study). If diffuse functions are required, either aug-pcseg-1 or 6-31++G** are recommended be used. Basis set errors are similar in magnitude to hybrid density functional approximation errors and thus are an important consideration when determining calculation accuracy.
\item \textbf{Triple zeta, doubly-polarised basis sets}: pcseg-2 (with  the M06-2X and $\omega$B97M-V functionals delivering similar performance in our limited study). If the system requires diffuse functions, def2-TZVPD is recommended. Basis set errors for triple-zeta basis sets are typically negligible compared to hybrid density functional approximation errors. 
    \item \textbf{On the use of 6-311G family:} Despite its name, polarised 6-311G should not be considered to be of triple valence zeta quality (this paper provides compelling quantitative evidence to support the very early results of \citet{grev_6-311g_1989} that have unfortunately often been ignored).
    \item \textbf{On including polarisation functions}: Incorporation of some polarisation functions is absolutely essential for high calculation quality, but some time savings without significant accuracy reduction can probably be obtained by reducing the level of polarisation on hydrogens. 
    \item \textbf{On including diffuse functions}: The use of diffuse functions for a particular molecular system needs to be carefully considered. When they are necessary (e.g. anions, strongly charged atoms), they result in a substantial improvement in accuracy that is observed clearly in mean and median absolute errors for the whole basis set. However, their use substantially increases computational time and often leads to significant SCF convergence issues and need to modify calculation thresholds. Ironically, these issues are particularly acute when diffuse functions are largely unnecessary to describe the system and don't result in significant accuracy improvement.  
\end{itemize}

Please note that this benchmarking study is focused on ground-state reaction energies of predominantly organic molecules. The generalisability of these conclusions to other calculation types needs to be considered carefully and there are certainly some known cases where they will not hold as the basis set requirements are very different. Most strikingly, for core properties such as NMR chemical shifts and coupling constants, the basis set requirements are very different and so a separate benchmarking study has been conducted \cite{IrelandSpecialisation2023}. For excited states, especially very highly excited states, diffuse functions are likely to be important as the electronic wavefunction is significantly larger than the ground state wavefunction. Diffuse functions are also known to be important when modelling dipole moments \cite{zapata2020computation} which depend on good descriptions of electron densities further from the nucleus. However, for properties that depend largely on valence chemistry, the recommendations can be robust; for example, the recent model chemistry benchmarking study for scaled harmonic vibrational frequencies \cite{zapata2023model} has similar conclusions for basis set performance as the current paper.

Within the constraints of ground-state reaction energy chemistry for main-group species, the main limitation of our study is the small benchmark data set and use of only 3 hybrid density functional methods. The useful follow-up studies here considering more levels of theory and a bigger set of molecular data are obvious, but there is one important caveat; designing a benchmarking set for evaluating basis set performance is subtly different from sets developed for evaluating density functional approximation performance. Most notably, many important basis sets are not defined across the periodic table. The other consideration is that diffuse augmented basis sets (i.e. those with low exponent diffuse functions) can be very challenging to converge for systems where they are unnecessary as there are significant linear dependency issues. We recommend, therefore, that a large basis set benchmarking data set should be divided into sub-categories based on whether they include molecules with uncommon elements and if diffuse functions are necessary. 

As a more general recommendation for benchmark data set design, we suggest that benchmark data sets and particularly diet versions of these data sets should consider typical calculation time in their construction; while of course a variety of molecular system sizes are important to include to ensure generality of the conclusions, it is unhelpful for a very small number of systems to dominate calculation time.

\begin{acknowledgement}

This research was undertaken with the assistance of resources from the National Computational Infrastructure (NCI Australia), an NCRIS enabled capability supported by the Australian Government.

The authors would also like to acknowledge A/Prof Frank Jensen for his insightful feedback on early drafts of this manuscript as well as A/Prof Gab Abramowitz for the idea of an average of normalised statistics (the performance statistic in this work).

\end{acknowledgement}

\begin{suppinfo}

The data underlying this study are available in the published article and its online supplementary material. 

We provide the xlxs file, with two data sheets, the first containing the raw energies for individual molecules, the second the individual reaction energies, both for all model chemistries considered in this manuscript. A read.me introduction sheet is included in the workbook describing each of the columns in the csv files. The optimised xyz files used can be downloaded from the GMTKN55 benchmark website.


\end{suppinfo}


\bibliography{references_basissets}

\clearpage

\end{document}